\documentclass[useAMS,usenatbib]{mn2e}

\title[CCD photometry of NGC 2425, Haffner~10 and Czernik~29]
{CCD photometry of distant open clusters\\
NGC 2425, Haffner~10 and Czernik~29}
\author[P. Pietrukowicz et al.]
  {P.~Pietrukowicz$^{1}$, J.~Kaluzny$^{1}$ and W.~Krzeminski$^{1,2}$\\
  $^1$Nicolaus Copernicus Astronomical Center,
     ul. Bartycka 18, 00-716 Warsaw, Poland
     (pietruk,jka@camk.edu.pl)\\
  $^2$Las Campanas Observatory, Casilla 601, La Serena, Chile
  (wojtek@lco.cl)\\ }

\date{Accepted .......
      Received .......;
      in original form ........}

\pagerange{\pageref{firstpage}--\pageref{lastpage}} \pubyear{2005}

\begin{document}

\maketitle

\label{firstpage}

\begin{abstract}
We present $BVI$ photometry for poorly known southern
hemisphere open clusters: NGC~2425, Haffner~10 and Czernik~29.
We have calculated the density profile and established the number
of stars in each cluster. The colour-magnitude diagrams of the
objects show a well-defined main sequence. However, the red giant clump is
present only in NGC~2425 and Haffner~10. For these two clusters
we estimated the age as $2.5 \pm 0.5$~Gyr assuming metallicity
of $Z=0.008$. The apparent distance moduli are in the
ranges $13.2<(m-M)_V<13.6$ and $14.3<(m-M)_V<14.7$,
while heliocentric distances are estimated
to be $2.9<d<3.8$~kpc and $3.1<d<4.3$~kpc, respectively
for NGC~2425 and Haffner~10. The angular separation of 2.4~deg
(150~pc at mean distance) may indicate a common origin of the
two clusters.

\end{abstract}


\section{Introduction} \label{s1}

The most recent catalogue by Dias et~al. (2002) presents
data on 1537 Galactic open clusters. These are excellent
objects with which to probe the structure and evolution
of the Galactic disc. Since old and intermediate-age
open clusters cover the entire lifetime of the disc,
they allow us to trace the history of chemical enrichment
and the formation of the different Galactic populations.
For example, Salaris, Weiss \& Percival (2004) determined the ages
of 71 old open clusters and concluded that the thin
and thick disc started to form at the same time.
Carraro et al. (2004) observed open clusters from the
outskirts of the disc and suggested that the outer part of the Galactic
disc underwent a completely different evolution compared
to the inner disc. Based on a sample of 39 open clusters,
Friel et~al. (2002) noted a slight correlation between metallicity
and age for clusters in the outer disc beyond 10 kpc.
Recently, intermediate-age open clusters were also found towards
the Galactic Centre direction (Carraro, M\'endez \& Costa 2005),
where star clusters do not survive for enough time due to
the higher-density environment.

The observations of the clusters presented in this  
paper were conducted as a part of a photometric survey
of a large sample of distant open clusters. The goal
of the project was an identification of the oldest open
clusters in order to obtain a lower limit on the age of the
Galactic disc (Kaluzny \& Rucinski 1995, see also references therein).
In this paper, we present for CCD photometry of three faint
open clusters: NGC 2425, Haffner~10 and Czernik~29.
The equatorial and galactic coordinates of the cluster  
centres are listed in Table~1.

\begin{table}
\centering
\begin{minipage}{200mm}
\caption{\small Equatorial and galactic coordinates of target
clusters}
{\small
\begin{tabular}{lcccc}
\hline
Name & RA(2000.0) & Dec(2000.0) & $l$ & $b$ \\
 & [h:m:s] & [$^{\circ}:'$] & [$^{\circ}$] & [$^{\circ}$] \\
\hline
NGC 2425 & 07:38:22 & -14:52.9 & 231.50 & 3.31 \\
Haffner~10 & 07:28:36 & -15:21.9 & 230.80 & 1.01 \\
Czernik~29 & 07:28:23 & -15:24.0 & 230.81 & 0.95 \\
\hline
\end{tabular}}
\end{minipage}
\end{table}


\section{Observations and reductions} \label{s2}

The observations were performed at the Las Campanas Observatory,
using the 1.0-m Swope telescope equipped with a Tektronix
$1024~\times~1024$ CCD camera. The field of view was about
$11 \farcm 9~\times~11 \farcm 9$ with a scale 0.695 arcsec/pixel.
The observations were conducted on two nights, Feb 20/21 and Feb 21/22,
1995. Two or three exposures of different length
were taken in each of the $BVI$ passbands. Preliminary processing of
the CCD frames was done with standard routines in the IRAF package.   
Both, dome and sky flatfield frames were obtained in each
filter. The average seeing was $1\farcs60$ and $2\farcs07$
on the first and second night, respectively.
We observed 49 standard stars from the three fields
(SA 98, Ru 149, PG 1323-086) listed by Landolt (1992), and 27 stars
from two fields (SA 98, Ru 149) during the two subsequent nights.
These standards were observed over airmasses from 1.07 to 1.76.  
The following relations between the instrumental (lower case letters)  
and the standard colours and magnitudes were adopted:
$$
v=2.561+V-0.019 \times (B-V)+0.15 \times X
\eqno(1)
$$
$$
b-v=0.214+0.931 \times (B-V)+0.12 \times X
\eqno(2)
$$
$$
v-i=-0.668+1.019 \times (V-I)+0.09 \times X
\eqno(3)
$$
where $X$ is an airmass. The instrumental photometry was extracted
with the DAOPHOT/ALLSTAR V2.0 (Stetson 1987) package. Aperture
photometry of standards was obtained with an aperture radius
of $8\farcs34$ arcsec (12 pixels). For stars from the cluster area we
obtained profile photometry. Appropriate aperture corrections
were derived preceding the transformation of instrumental photometry
to the standard system.
  
We applied the following procedure to a set of images
obtained for a given cluster. First, objects with unusually large
errors of photometry, i.e. with large values of CHI and SHARP
parameters, returned by DAOPHOT, were rejected. Also very bright, 
overexposed stars were removed from further analysis. In practice,
the percentage of rejected stars ranged for a given frame   
from 5 to 15 percent. Second, the coordinates of all objects
were transformed to a common pixel grid defined by the reference image
(always the longest exposure in the $V$ filter). We than corrected  
the photometry for the zero-point offset in each filter and
created a master list of all objects. The instrumental magnitudes
were calculated as weighted averages of magnitudes measured
on individual frames.


\section{NGC 2425} \label{s3}

The open cluster NGC 2425 (C 0736-147) lacks any published
study. We observed it on 1995 Feb 20/21.
Details of the observations are presented in Table 2.

\begin{table}
\centering
\begin{minipage}{200mm}
\caption{\small Journal of observations of NGC 2425}
{\small
\begin{tabular}{lccccccc}
\hline
UT Date & Filter & Exp. & Airmass & Seeing \\
Feb 1995 & & [sec] & & [arcsec] & \\
\hline
21.092 & $I$ &   5 & 1.031 & 1.70 \\
21.094 & $I$ &  15 & 1.031 & 1.47 \\
21.098 & $I$ & 100 & 1.031 & 1.56 \\
21.101 & $V$ &  15 & 1.032 & 1.70 \\
21.102 & $V$ & 100 & 1.032 & 1.63 \\
21.104 & $V$ & 300 & 1.032 & 1.51 \\
21.110 & $B$ &  30 & 1.034 & 1.64 \\
21.111 & $B$ & 150 & 1.035 & 1.51 \\
21.114 & $B$ & 500 & 1.037 & 1.64 \\
\hline
\end{tabular}}
\end{minipage}
\end{table}

The cluster centre was found by calculating the density centre
for stars inside a circle of radius of 150 pixels, using an
iterative procedure similar to that described by Mateo \& Hodge (1986).
In Fig.~1 we show the density profile for all stars brighter
than $V=21.9$. The average stellar density was
calculated in successive 27.8 arcsec (40 pixels) wide annuli
around the cluster centre. The resulting density
profile is rather noisy due to small number statistics.
The smooth  
solid line represents a fit by the King (1962) profile:
$$
f(r) = \frac{f_0}{1+(r/r_c)^2} + f_b
\eqno(4)
$$
where $f_0$ is the central density, $r_c$ is the radius of the cluster
core and $f_b$ is the background density.
For NGC 2425 we found as follows:
$f_0=0.0056 \pm 0.0008$ stars per arcsec$^2$,   
$r_c=154 \pm 41$ arcsec and $f_b=0.0039 \pm 0.0004$ stars/arcsec$^2$.
We estimate that the cluster contains about 850 stars with $V<21.9$.

\begin{figure}
\vspace{5.9cm}
\includegraphics{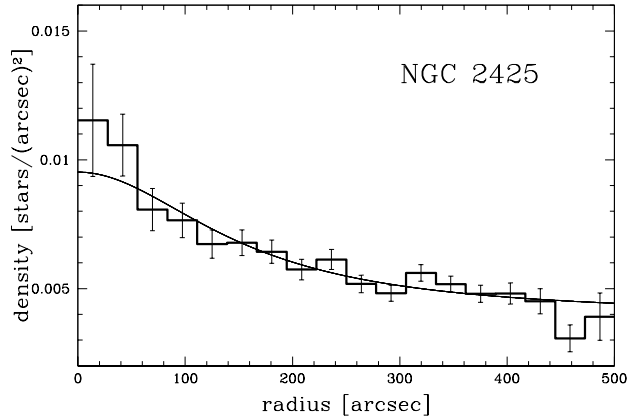}
\caption{\small Surface density distribution of stars with $V<21.9$
from NGC 2425 field. The King (1962) profile is fit to the data
and presented as the smooth solid line.}  
\label{fig1}
\end{figure}

\begin{figure}
\vspace{5.9cm}
\includegraphics{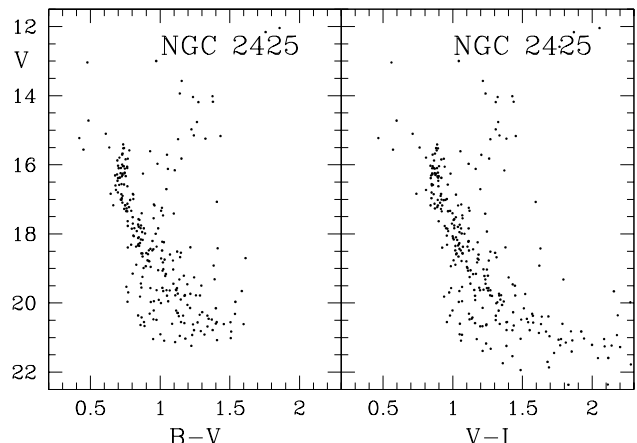}   
\caption{\small CMDs for the stars located within radius
$r<0.7~r_c\approx108\arcsec$ from the centre of NGC 2425}
\label{fig2}
\end{figure}

Colour-magnitude diagrams for the field of NGC 2425
are shown in two panels of Fig.~2. We plot only the innermost
stars to get some structures, in particular the main sequence
and red clump area. We can conclude that the morphology
of CMDs for NGC 2425 is typical of intermediate-age
open clusters. We note the presence of the red giant branch clump  
at $V \approx 14.1$, $B-V \approx 1.37$ and $V-I \approx 1.31$. 
One can also distinguish the main sequence turn-off point
at $V \approx 16.5$, $B-V \approx 0.68$ and $V-I \approx 0.84$.
We suspect that there are candidates for blue stragglers among
stars above the turn-off.

Using theoretical isochrones published by Girardi et~al.
(2000), we are able to estimate the basic parameters of the cluster.  
We fit isochrones with two different chemical compositions:
$(Z,Y)=(0.008,0.25)$ and $(0.019,0.273)$. In Fig.~3 we show
CMDs with superimposed isochrones of age of 2.5~Gyr and lower metal content,
$Z=0.008$. The shape of the main sequence for $16<V<20$ is well reproduced,
while the blue and red hooks are not clearly seen. Comparing
other isochrones to the data we estimated the uncertainty of the age
as 0.5~Gyr. We derived, by shifting the isochrones, the apparent distance
modulus of $(m-M)_V=13.5$, the reddenings of $E(B-V)=0.29$ and
$E(V-I)=0.34$. These values are lower than the upper limits of
reddening $E(B-V)=0.47$ and $E(V-I)=0.60$ for $(l,b)=(231.5,+3.3)$
extracted from the maps of Schlegel, Finkbeiner \& Davis (1998). We assume
the error of the distance modulus as 0.1~mag, and the error
of the reddening $E(B-V)$ as 0.05~mag.

Fig.~4 presents the CMDs of NGC 2425 with superimposed isochrones
of the same age, as in Fig.~3, but with solar metal abundance,
$Z=0.019$. The fit is worse just above the turn-off point while the 
RGB branch is reproduced quite well. In this case we established:
$(m-M)_V=13.3 \pm 0.1$, $E(B-V)=0.20 \pm 0.05$, $E(V-I)=0.26 \pm 0.05$.
Adopting $R_V=3.2$ and taking into account the results for both
metallicites we estimated the minimum value of the
heliocentric distance of the cluster as $d_{min}=2.9$~kpc  
and the maximum value $d_{max}=3.8$~kpc.

\begin{figure}
\vspace{5.9cm}
\includegraphics{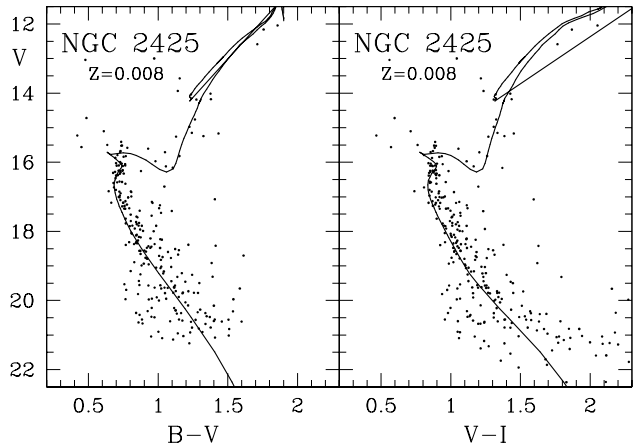}
\caption{\small Left panel: $V/B-V$ diagram
for the cluster NGC 2425, as compared to Girardi et~al. (2000)
isochrone of age $2.5 \times 10^9$~yr and metallicity $Z=0.008$
(solid line). The fit was obtained by adopting a distance
modulus of $(m-M)_V=13.5$ and reddening of $E(B-V)=0.29$.
Right panel: field-star corrected $V/V-I$ diagram
for NGC 2425 cluster, as compared to Girardi et~al. (2000)
isochrone of age $2.5 \times 10^9$~yr for the metallicity $Z=0.008$
(solid line). The fit was obtained by adopting a distance
modulus of $(m-M)_V=13.5$ and reddening of $E(V-I)=0.34$.}
\label{fig3}
\end{figure}

\begin{figure}
\vspace{5.9cm}
\includegraphics{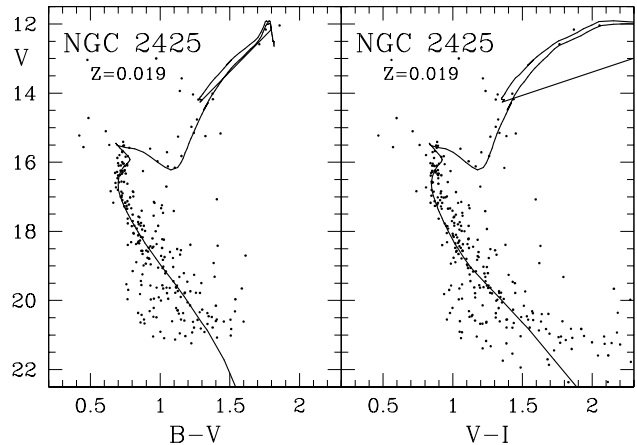}    
\caption{\small The same as Fig. 3, but for $Z=0.019$.
The fit in the left panel was obtained by adopting a distance
modulus of $(m-M)_V=13.3$ and reddening of $E(B-V)=0.20$,
whereas the fit in the right panel by adopting
$(m-M)_V=13.3$ and $E(V-I)=0.26$.}
\label{fig4}
\end{figure}


\section{Haffner~10 and Czernik~29} \label{s4}

The clusters Haffner~10 (OCl 594, C 0726-152) and Czernik~29
(OCl 595, C 0726-153) were identified by Haffner (1957)
and Czernik (1966), respectively. Later Fitzgerald \& Moffat (1980)
studied the clusters based on photographic $UBV$ plates.
However, due to rather low photometric
limits ($B=18$, $V=15.1$) they underestimated the number
of stars in the clusters ($68\pm12$ in Haffner~10 and $56\pm11$ in
Czernik~29 based on $B$ filter data) and the age of Haffner~10 (0.2 Gyr).
We observed these clusters on 1995 Feb 20/21
and 21/22. The list of used CCD frames is given in Table 3.
The angular separation between the clusters is only 3 \farcm 8
and both of them were embraced within each frame.

\begin{table}
\centering
\begin{minipage}{200mm}
\caption{\small Journal of observations of Haffner~10}
{\small
\begin{tabular}{lccccccc}
\hline
UT Date & Filter & Exp. & Airmass & Seeing \\
Feb 1995 & & [sec] & & [arcsec] & \\
\hline
21.127 & $I$ &  10 & 1.054 & 1.37 \\
21.129 & $I$ &  60 & 1.057 & 1.81 \\
21.130 & $I$ & 180 & 1.059 & 1.50 \\
21.135 & $V$ &  20 & 1.067 & 1.43 \\
21.136 & $V$ & 100 & 1.069 & 1.63 \\
21.138 & $V$ & 360 & 1.073 & 1.63 \\
21.144 & $B$ &  40 & 1.084 & 1.69 \\
21.146 & $B$ & 150 & 1.089 & 1.63 \\
21.149 & $B$ & 500 & 1.093 & 1.72 \\
22.056 & $V$ &  20 & 1.046 & 2.36 \\
22.058 & $B$ &  40 & 1.043 & 2.30 \\
22.060 & $I$ &  10 & 1.041 & 1.86 \\
22.062 & $I$ & 300 & 1.040 & 2.03 \\
22.067 & $V$ & 600 & 1.036 & 2.00 \\
22.077 & $B$ & 900 & 1.031 & 1.88 \\
\hline
\end{tabular}}
\end{minipage}
\end{table}

As for previous cluster, for Haffner~10 and Czernik~29
we present the density histograms (Fig.~5). The histograms
include stars with $V<21.7$. We determined the centres
of the clusters and calculated the average stellar density
in successive 27.8 arcsec (40 pixels) annuli around the centre,
excluding regions within the $1 \farcm 9$ area around
the neighbouring cluster. This radius is equal to half
the angular distance between the clusters,
which are of comparable size (from Fitzgerald \& Moffat 1980).

For Haffner~10 the best fit of King's profile results
in the following coefficients:
the central density $f_0=0.0123 \pm 0.0020$ stars/arcsec$^2$,
the radius of the cluster core $r_c=65 \pm 11$ arcsec and the
background density $f_b=0.0051 \pm 0.0002$ stars/arcsec$^2$.
We established the number of cluster stars with magnitude $V<21.7$
to be approximately 600 objects. This was determined by adopting
the above value of $f_b$. For Czernik~29 we found:
the central density $f_0=0.0066 \pm 0.0013$ stars/arcsec$^2$,
the radius of the cluster core $r_c=78 \pm 17$ arcsec and the
background density $f_b=0.0053 \pm 0.0002$ stars/arcsec$^2$.
The number of cluster stars with magnitude $V<21.7$
is approximately 420 objects.

\begin{figure}
\vspace{8.3cm}
\includegraphics{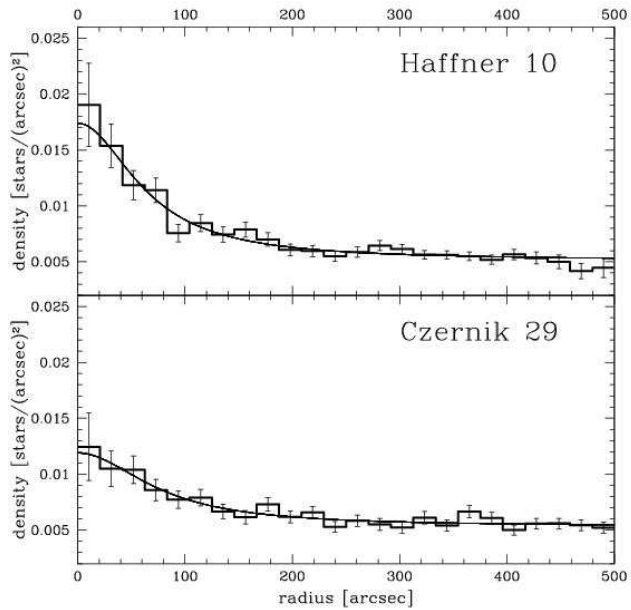}
\caption{\small Surface density distributions of stars with $V<21.7$
from the field of Haffner~10 and Czernik~29}
\label{fig5}
\end{figure}

In Fig.~6 we present $V/B-V$ and $V/V-I$ colour-magnitude diagrams
for Haffner~10. The morphology of CMDs is typical
for intermediate-age open clusters.
Interestingly, the red clump is represented by a tilted
branch. Its width reaches 0.20 in the $B-V$ and 0.22 in the $V-I$.
The branch is elongated in the direction parallel to the standard
reddening vector. It means that there is significant
differencial absorption in the cluster direction.
The blue turn-off point is located at $V \approx 17.3$,
$B-V \approx 0.92$ and $V-I \approx 1.18$, while
the blue end of the red clump is located at $V \approx 14.8$, 
$B-V \approx 1.41$ and $V-I \approx 1.58$.

\begin{figure}
\vspace{5.9cm}
\includegraphics{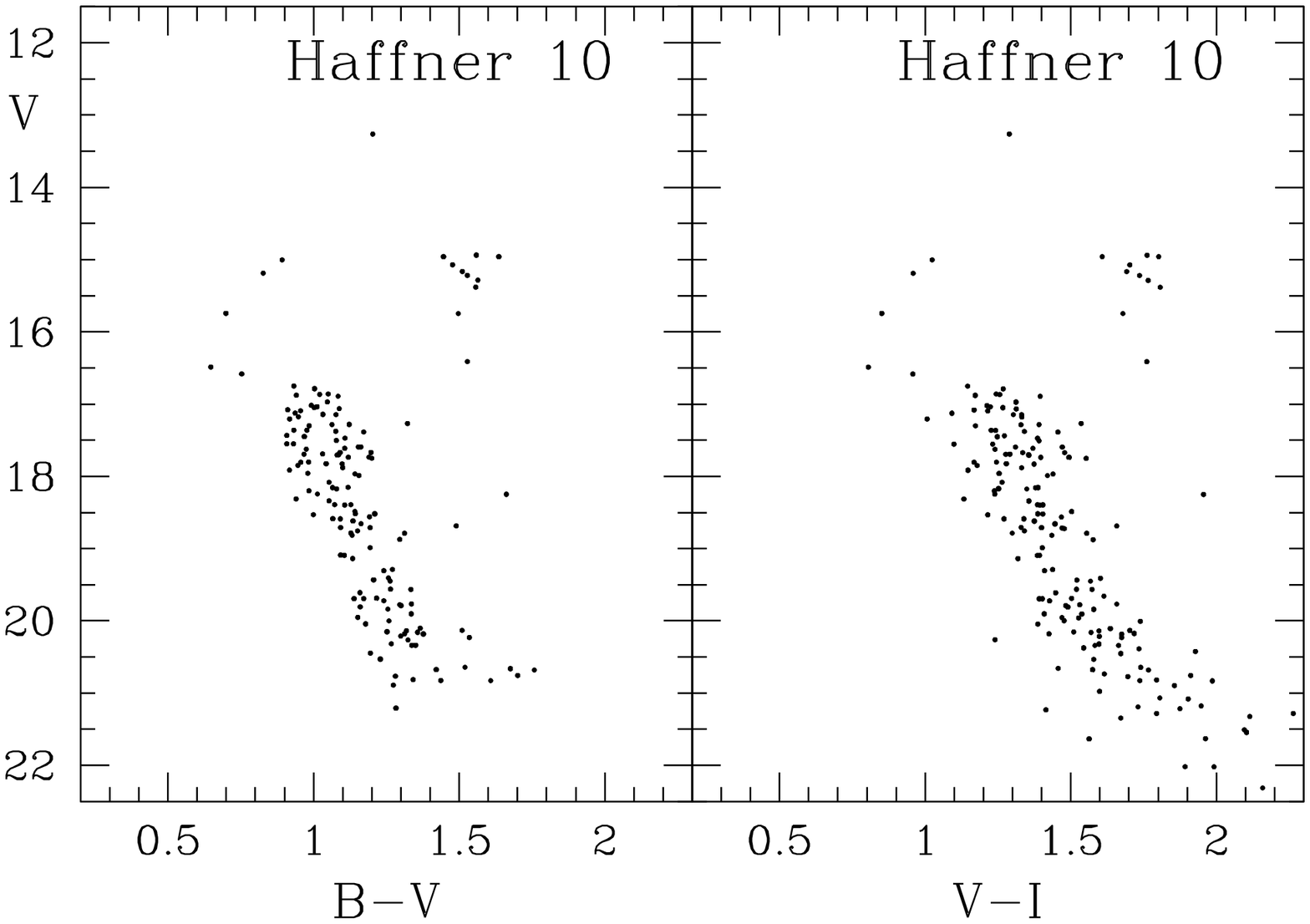}
\caption{\small CMDs for the stars located within radius 
$r<1.2~r_c\approx78\arcsec$ from the centre of Haffner~10}
\label{fig6}
\end{figure}

Using a set of theoretical isochrones published by
Girardi et~al. 2000 we estimated some cluster parameters,
in particular the age and the heliocentric distance.
As in the case of NGC 2425, we adopted isochrones for
two different metallicites. Figures 7 and 8 show
CMDs with superimposed isochrones of 2.5~Gyr for metallicity
$Z=0.008$ and of 2.0~Gyr for $Z=0.019$, respectively. 
The fit for lower metal content seems to be more precise.
However we note that it is difficult to establish
the location of the blue and red hooks.   
Trying to improve the fit and comparing other isochrones we estimated
the error of the age as 0.5~Gyr. By shifting the isochrones for both
metallicities, we obtained the value of the apparent distance modulus
$(m-M)_V$ between 14.3 and 14.7, the reddenings $0.41<E(B-V)<0.64$
and $0.58<E(V-I)<0.78$. This is in agreement with the upper value
of the interstellar reddening, $E(B-V)=0.89$ and $E(V-I)=1.14$,
for $(l,b)=(230.8,+1.0)$ derived from maps presented by
Schlegel, Finkbeiner \& Davis (1998). We estimated that the heliocentric
distance $d$ to Haffner~10 is in the range between 3.1 and 4.3 kpc.

Figure 9 shows $V/B-V$ and $V/V-I$ colour-magnitude diagrams
for Czernik~29. The cluster lacks the red giant clump, which makes
estimation of its age and distance difficult. The main sequence
has brighter stars than the main sequence of Haffner~10,
therefore Czernik~29 is either younger or/and has a lower distance
than Haffner~10.

\begin{figure}
\vspace{5.9cm}
\includegraphics{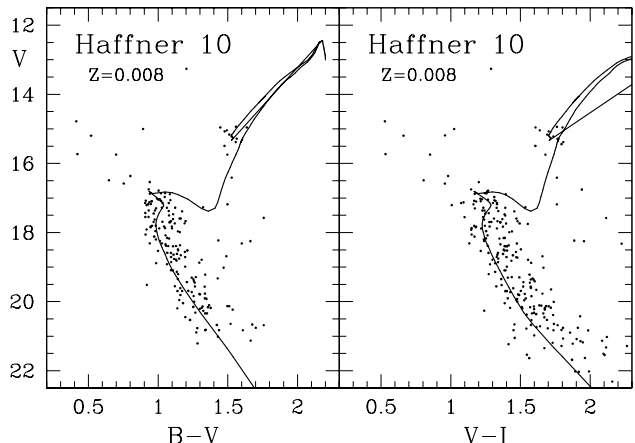}
\caption{\small Left panel: $V/B-V$ diagram
for the cluster Haffner~10, as compared to Girardi et~al. (2000)
isochrone of age $2.5 \times 10^9$~yr for the metallicity $Z=0.008$
(solid line). The fit was obtained by adopting a distance
modulus of $(m-M)_V=14.6$ and reddening of $E(B-V)=0.59$.
Right panel: field-star corrected $V/V-I$ diagram
for the Haffner~10 cluster, as compared to Girardi et~al. (2000)
isochrone of age $2.5 \times 10^9$~yr for the metallicity $Z=0.008$
(solid line). The fit was obtained by adopting a distance
modulus of $(m-M)_V=14.6$ and reddening of $E(V-I)=0.73$.}
\label{fig7}
\end{figure}

\begin{figure}
\vspace{5.9cm}
\includegraphics{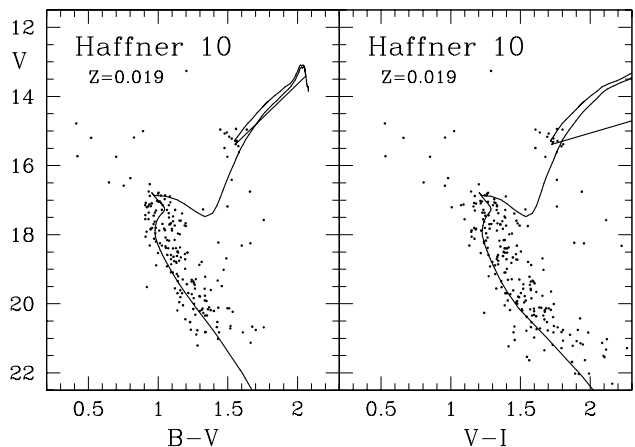}
\caption{\small CMDs for the cluster Haffner~10 with
superimposed isochrones of age $2.8 \times 10^9$~yr
for the metallicity $Z=0.019$.
The fit in the left panel was obtained by adopting a distance
modulus of $(m-M)_V=14.4$ and reddening of $E(B-V)=0.46$,
whereas the fit in the right panel by adopting
$(m-M)_V=14.4$ and $E(V-I)=0.63$.}
\label{fig8}
\end{figure}

\begin{figure}
\vspace{5.9cm}
\includegraphics{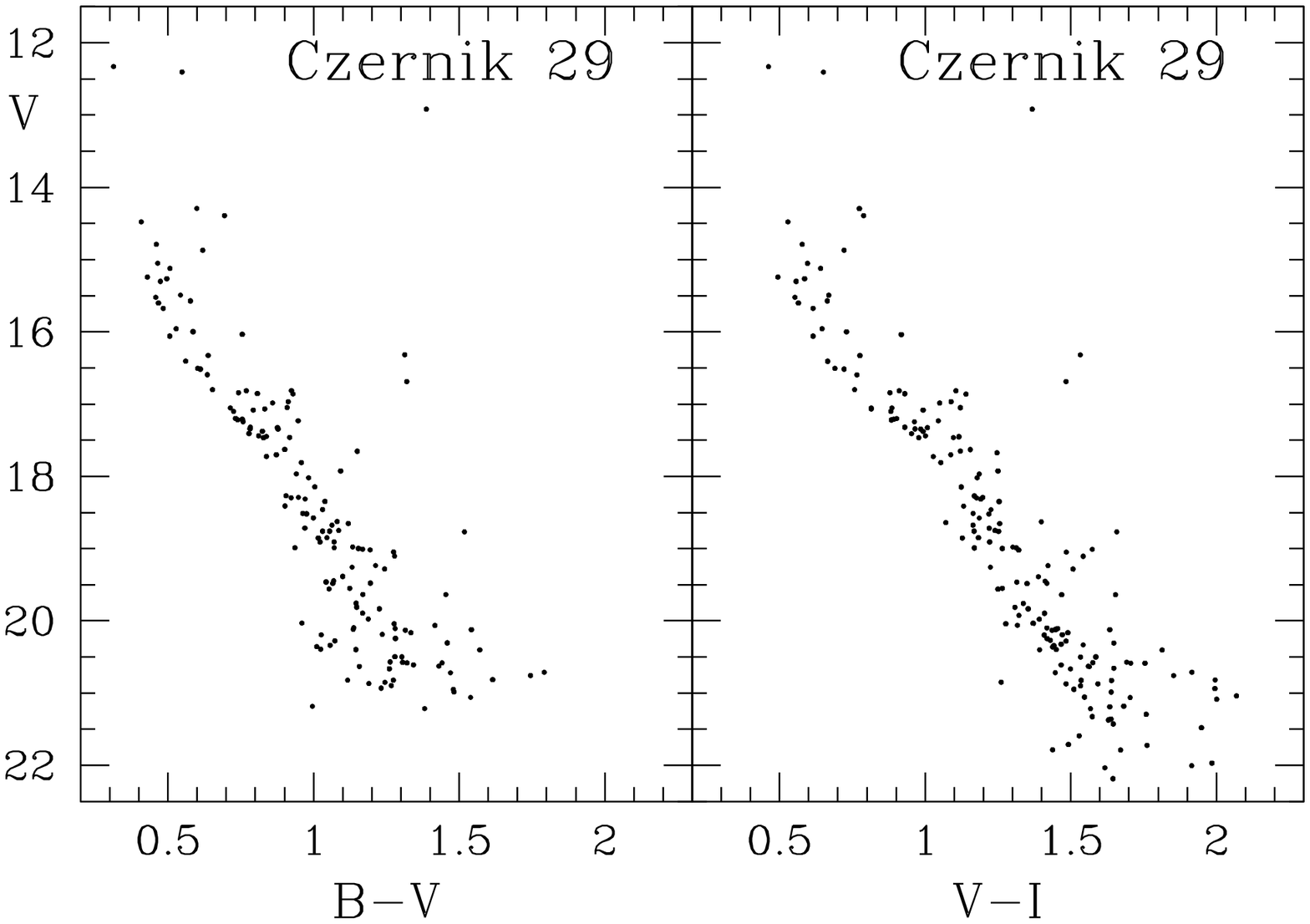}
\caption{\small CMDs for the stars located within radius 
$r<1.2~r_c\approx94\arcsec$ from the centre of Czernik~29}
\label{fig8}  
\end{figure}


\section{Summary} \label{s5}

\begin{table*}
\begin{center}
\caption{\small Observed and determined parameters of analysed 
open clusters}

\vspace{0.4cm}

\begin{tabular}{lcccccccc}
\hline
Name & $r_c$ & N & $(m-M)_V$ & $E(B-V)$ & $d$ & Age \\
    & [arcsec] & & [mag] & [mag] & [kpc] & [Gyr] \\
\hline
NGC 2425    & 155 & 850 & 13.2-13.6 & 0.15-0.34
 & 2.9-3.8 & $2.5 \pm 0.5$ \\
Haffner~10  & 65 & 600 & 14.3-14.7 & 0.41-0.64
 & 3.1-4.3 & $2.5 \pm 0.5$ \\
Czernik~29  & 78 & 420 & - & - & - & - \\
\hline
\end{tabular}
\end{center}
\end{table*}

We have presented $BVI$ photometry for three poorly studied open
clusters from the southern hemisphere. For each cluster we calculated
surface density profile and established the probable number of cluster
members. The analysis of the derived colour-magnitude diagrams
allowed us to estimate basic parameters like the age and distance.
The results are summarized in Table 4. We found that the clusters
NGC 2425 and Haffner~10, are intermediate-age open clusters.
Probably due to the relatively large number of star members,
the clusters still constitute physical systems.

We notice that the age as well as the heliocentric distance
estimations of the two clusters are very similar, though
the extinctions are quite different.
Interestingly, the angular separation between the clusters is only
$2 \fdg 4$, which gives an approximate linear seperation of  
150~pc, assuming an average distance of 3.65~kpc.
This is about 50 and 100 times larger than the cluster
core radius of NGC 2425 and Haffner~10, respectively.
We should note that the seperation is larger than for other double
systems of clusters presented in the literature
(Subramaniam et~al. 1995, Pietrzy\'nski \& Udalski 2000).
However, there is a possibility that the two clusters
constituted a pair of open clusters in the past.

Based on the comparison of the CMDs with the theoretical
isochrones we were not able to firmly establish the metallicity
of either of the clusters. We may only suggest that the metal abundance
is comparable for both clusters, which may indicate a common origin.
A spectroscopic determination of the metallicity and radial
velocities would help to verify this hypothesis.


\section*{Acknowledgments}

We would like to thank Greg Stachowski for remarks on
the draft version of this paper.
PP was supported by the grant {\bf 1~P03D 024 26} from the
State Committee for Scientific Research, Poland.


\begin{thebibliography}{99}
\bibitem[Carraro et al. 2004]{carr04} Carraro G., Bresolin F., Villanova S.,
   Matteucci F., Patat F., Romaniello M., 2004, AJ, 128, 1676
\bibitem[Carraro, M\'endez \& Costa, 2005]{carr05} Carraro G., M\'endez A.,
   Costa E., 2005, MNRAS, 356, 647
\bibitem[Czernik, 1966]{czer66} Czernik M., 1966, AcA, 16, 93
\bibitem[Dias et al. 2002]{dias02} Dias W. S., Alessi B.~S., Moitinho A.,
   L\'epine J.~R.~D., 2002, A\&A, 389, 871
\bibitem[Fitzgerald \& Moffat, 1980]{fitz80} Fitzgerald M.~P.,
   Moffat A.~F.~J., 1980, PASP, 92, 489
\bibitem[Friel et al. 2002]{fri02} Friel E.~D., Janes K.~A., Tavarez M.,
   Scott J., Katsanis R., Lotz J., Hong L., Miller N., 2002, AJ, 124, 2693
\bibitem[Girardi et al. 2000]{gir00} Girardi L., Bressan A., Bertelli G.,
   Chiosi C., 2000, A\&AS, 141, 371
\bibitem[Haffner, 1957]{haff57} Haffner H., 1957, ZAp, 43, 89
\bibitem[Landolt, 1992]{lan92} Landolt A.~U., 1992, ApJ, 104, 340
\bibitem[Kaluzny \& Rucinski, 1995]{kal95} Kaluzny J., Rucinski S.~M.,
   1995, A\&AS, 114, 1
\bibitem[King, 1962]{king62} King I., 1962, AJ, 67, 471
\bibitem[Mateo \& Hodge, 1986]{mat86} Mateo M.~M., Hodge P.,
   1986, ApJS, 60, 833
\bibitem[Pietrzy\'nski \& Udalski, 2000]{pie00} Pietrzy\'nski G.,
   Udalski A., 2000, AcA, 50, 355
\bibitem[Salaris, Weiss \& Percival, 2004]{sal04} Salaris M., Weiss A.,
   Percival S.~M., 2004, A\&A, 414, 163
\bibitem[Schlegel, Finkbeiner \& Davis, 1998]{sch98} Schlegel D.~J.,
   Finkbeiner D.~P., Davis M., 1998, ApJ, 500, 525
\bibitem[Stetson, 1987]{stet87} Stetson P.~B., 1987, PASP, 99, 191
\bibitem[Subramaniam et al. 1995]{sub95} Subramaniam A., Gorti U.,
   Sagar R., Bhatt H.~C., 1995, A\&A, 302, 86
\end{thebibliography}
\end{document}